\documentstyle[11pt,amssymb,epsf,cite]{article}

\makeatletter
\@addtoreset{equation}{section}
\makeatother

\def\ben{\begin{equation}}
\def\een{\end{equation}}

  \let\n=\nu

\let\C=\Chi

 \def\bd{\begin{document}} \def\ed{\end{document}}
\def\ds{\documentstyle} \let\fr=\frac \let\bl=\bigl \let\br=\bigr
\let\Br=\Bigr \let\Bl=\Bigl
\let\bm=\bibitem
\let\na=\nabla
\let\pa=\partial \let\ov=\overline
\newcommand{\be}{\begin{equation}}
\newcommand{\ee}{\end{equation}}
\def\ba{\begin{array}}
\def\ea{\end{array}}
\def\ft#1#2{{\textstyle{{\scriptstyle #1}\over {\scriptstyle #2}}}}
\def\fft#1#2{{#1 \over #2}}
\def\del{\partial}
\def\vp{\varphi}
\def\sst#1{{\scriptscriptstyle #1}}
\def\oneone{\rlap 1\mkern4mu{\rm l}}
\def\td{\tilde}
\def\wtd{\widetilde}
\def\ie{\rm i.e.\ }
\def\dalemb#1#2{{\vbox{\hrule height .#2pt
        \hbox{\vrule width.#2pt height#1pt \kern#1pt
                \vrule width.#2pt}
        \hrule height.#2pt}}}
\def\square{\mathord{\dalemb{6.8}{7}\hbox{\hskip1pt}}}
\newcommand{\ho}[1]{$\, ^{#1}$}
\newcommand{\hoch}[1]{$\, ^{#1}$}
\newcommand{\bea}{\begin{eqnarray}}
\newcommand{\eea}{\end{eqnarray}}
\newcommand{\ra}{\rightarrow}
\newcommand{\lra}{\longrightarrow}
\newcommand{\Lra}{\Leftrightarrow}
\newcommand{\ap}{\alpha^\prime}
\newcommand{\bp}{\tilde \beta^\prime}
\newcommand{\tr}{{\rm tr} }
\newcommand{\Tr}{{\rm Tr} }
\def\0{{\sst{(0)}}}
\def\1{{\sst{(1)}}}
\def\2{{\sst{(2)}}}
\def\3{{\sst{(3)}}}
\def\4{{\sst{(4)}}}
\def\5{{\sst{(5)}}}
\def\6{{\sst{(6)}}}
\def\7{{\sst{(7)}}}
\def\8{{\sst{(8)}}}
\def\n{{\sst{(n)}}}
\def\cA{{{\cal A}}}
\def\cF{{{\cal F}}}
\def\tV{\widetilde V}
\def\tW{\widetilde W}
\def\tH{\widetilde H}
\def\tE{\widetilde E}
\def\tF{\widetilde F}
\def\tA{\widetilde A}
\def\im{{{\rm i}}}
\def\tY{{{\wtd Y}}}
\def\ep{{\epsilon}}
\def\vep{{\varepsilon}}
\def\R{\rlap{\rm I}\mkern3mu{\rm R}}
\def\bD{{{\bar D}}}

\def\R{\rlap{\rm I}\mkern3mu{\rm R}}
\def\bD{{{\bar D}}}
\def\R{{{\Bbb R}}}
\def\C{{{\Bbb C}}}
\def\H{{{\Bbb H}}}
\def\CP{{{\Bbb C}{\Bbb P}}}
\def\RP{{{\Bbb R}{\Bbb P}}}
\def\Z{{{\Bbb Z}}}
\def\bA{{{\Bbb A}}}
\def\bB{{{\Bbb B}}}
\def\bC{{{\Bbb C}}}
\def\bR{{{\Bbb R}}}
\def\bD{{{\Bbb D}}}
\def\bE{{{\Bbb E}}}
\def\bZ{{{\Bbb Z}}}
\def\cD{{{\cal D}}}
\def\Re{{{\frak{Re}}}}
\def\Im{{{\frak{Im}}}}
\def\cosec{{\,\hbox{cosec}\,}}
\def\Gm{{\Gamma_{\!\! -}}}
\def\Gp{{\Gamma_{\!\! +}}}

\def\stan{{standard }}
\def\nonstan{{supernumerary }}

\def\cosech{{\hbox{cosech}}}

\def\etcyc{{\hbox{and cyclic}}}
\def\btheta{{\bar\theta}}

\newcommand{\tamphys}{\it Center for Theoretical Physics,
Texas A\&M University, College Station, TX 77843, USA}
\newcommand{\umich}{\it Michigan Center for Theoretical Physics,
University of Michigan\\ Ann Arbor, MI 48109, USA}
\newcommand{\upenn}{\it Department of Physics and Astronomy,\\
University of Pennsylvania, Philadelphia,  PA 19104, USA}
\newcommand{\SISSA}{\it  SISSA-ISAS and INFN, Sezione di Trieste\\
Via Beirut 2-4, I-34013, Trieste, Italy}

\newcommand{\mitchell}{\it George P. \& Cynthia W.
Mitchell Institute for Fundamental Physics,\\
Texas A\&M University, College Station, TX 77843-4242, USA}

\newcommand{\newton}{\it Isaac Newton Institute for Mathematical Sciences,\\
0 Clarkson Road,  University of Cambridge,
Cambridge CB3 0EH, UK}

\newcommand{\ihp}{\it Institut Henri Poincar\'e\\
  11 rue Pierre et Marie Curie, F 75231 Paris Cedex 05}

\newcommand{\damtp}{\it DAMTP, Centre for Mathematical Sciences,
 Cambridge University,\\  Wilberforce Road, Cambridge CB3 OWA, UK}

\newcommand{\itp}{\it Institute for Theoretical Physics, University of
California\\ Santa Barbara, CA 93106, USA}

\newcommand{\istanbul}{\it Department of Mathematics,  Bo{\u g}azi{\c c}i
University, Bebek, Istanbul 34342, Turkey.}

\newcommand{\auth}{ R. G\"uven\footnote{Present address: Department of Physics,
I{\c s}$\imath$k University, Kumbaba Mevkii, {\c S}ile, Istanbul
34980, Turkey}}

\begin{document}
\begin{flushright}

\end{flushright}
\vspace{12pt}\vspace{12pt}\vspace{12pt}\vspace{14pt}
\begin{center}

{\Large {\bf The Conformal Penrose Limit and the Resolution of the
pp-curvature Singularities}} \vspace{14pt} \vspace{12pt}

\auth

\vspace{7pt}

\vspace{7pt}
\istanbul

\vspace{7pt}

\vspace{14pt} \vspace{14pt}\vspace{14pt}\vspace{12pt}
\vspace{14pt}

\underline{ABSTRACT}
\end{center}

    We consider the exact solutions of the supergravity theories
in various dimensions in which the space-time has the form $M_{d}
\times S^{D-d}$ where  $M_{d}$ is an Einstein space admitting a
conformal Killing vector and $S^{D-d}$ is a sphere of an
appropriate dimension. We show that, if the cosmological constant
of $M_{d}$ is negative and the conformal Killing vector is
space-like, then such solutions will have a conformal Penrose
limit: $M^{(0)}_{d}\times S^{D-d}$ where $M^{(0)}_{d}$ is a
generalized $d$-dimensional AdS plane wave. We study the
properties of the limiting solutions and find that $M^{(0)}_{d}$
has 1/4 supersymmetry as well as a Virasoro symmetry. We also
describe how the pp-curvature singularity of $M^{(0)}_{d}$ is
resolved in the particular case of the $D6$-branes of $D=10$ type
IIA supergravity theory. This distinguished case provides an
interesting generalization of the plane waves in $D=11$
supergravity theory and suggests a  duality between the $ SU(2)$
gauged $d=8$ supergravity theory of Salam and Sezgin on
$M^{(0)}_{8}$ and the $d=7$ ungauged supergravity theory on its
pp-wave boundary.

\pagebreak
\setcounter{page}{1}

\newpage

\section{Introduction}

   A considerable portion of our present
understanding of the properties of the string theories and of
M-theory is based on  two different notions of limits. The first
of these is the notion of the near-horizon limit of a BPS
$p$-brane\cite{gt}. In this limit one usually ends up with a
$AdS_{p+2} \times S^{D-p-2} $ type of geometry which is a product
of an anti de Sitter ($AdS$) space and a  sphere $S$ of
appropriate dimensions that add up to the total dimension $D$. The
second notion is that of the Penrose limit\cite{pen} which
implies, when suitably generalized to supergravity
theories\cite{guv}, that any supergravity solution has a plane
wave solution as a limit. Recently, it was found that the Penrose
limits of $AdS_{p+2} \times S^{D-p-2} $ type of geometries are the
maximally supersymmetric plane waves \cite{bla} and these provide
in D=10 a convenient setting for the quantization of superstrings
with non-trivial Ramond-Ramond (RR) fields \cite{mat}. Remarkably,
a string theory counterpart of the Penrose limit also
exists\cite{mald} and these developments led to new insights about
the AdS/CFT correspondence in a regime which is beyond the
supergravity approximation.

    In the light of these developments it is of interest to
explore the possible generalizations of the Penrose limit together
with the neighboring geometries of  the $AdS_{p+2} \times
S^{D-p-2} $ space-times. The purpose of the present paper is to
carry out such a program which is based on the existence of
conformal Killing vectors. We shall be interested primarily with
the space-like conformal Killing vectors and although most of our
considerations will be valid for all $D > 3$, we shall be mainly
concerned with the $D=10$ and $D=11$ solutions.

     In four dimensions $AdS_4$ is known to be the unique space-time
which admits a space-like conformal Killing vector (CKV) and a
negative cosmological constant \cite{gar}. It is, however,
possible to construct a two-parameter family of Einstein spaces
which  admits a space-like CKV as well as a negative cosmological
constant when one moves to higher dimensions. These geometries
have a warped product structure and give rise to a class of exact
solutions of the supergravity theories which are of the
Freund-Rubin type .  We shall be concerned with such classes of
solutions that include the $AdS_{p+2} \times S^{D-p-2} $
space-times as special cases.

 We  shall start by considering supergravity solutions in the dual frame
assuming that the space-time is of  the form $M_{p+2} \times
S^{D-p-2}$ where  $M_{p+2}$ is an Einstein space admitting a CKV.
These solutions will have in general  non-zero $(D-p-2)$-form
fluxes and possess dilaton fields that may play the role of
conformal Killing potentials on $M_{p+2}$. We shall show, by
studying the null geodesic congruences of such space-times and
employing the limiting procedure of\cite{guv2}, that these
solutions  have limits that are of the form $M^{(0)}_{p+2}\times
S^{D-p-2}$ where $M^{(0)}_{p+2}$ is a generalized
$(p+2)$-dimensional $AdS$ plane wave\cite{guv2} which reduces to
the non-linear version of the Randall-Sundrum zero mode\cite{chg}
when the CKV is hypersurface orthogonal. We shall also see that
these limits exist only when the CKV of $M_{p+2}$ is space-like
and the cosmological constant of $M_{p+2}$ is negative.

As in the case of the Kaigorodov space-time\cite{pope}, AdS plane
waves are known to preserve $1/4$ supersymmetries and possess a
Virasoro symmetry\cite{ban}. These space-times have been
extensively studied\cite{po}\cite{bre} and were found to suffer
from pp-curvature singularities. In this paper we shall show that
all of these properties are shared by the general limiting
solution $M^{(0)}_{p+2}$. We shall also describe how the
pp-curvature singularity is resolved in the case of $p=6$, $D=10$
type IIA supergravity by lifting up the solution to the $D=11$
supergravity theory. This particular case will be seen to play a
distinguished role because the CKV will be forced to be space-like
by the field equations themselves and will lead us to an
interesting generalization of plane waves in $D=11$ as well as to
a duality conjecture between  the  $SU(2)$ gauged $d = 8$
supergravity theory of Salam and Sezgin\cite{ss} on the limiting
background $M^{(0)}_{8}$  and the ungauged supergravity theory on
its $d=7$ pp-wave boundary.

\section{Dual Frame Products and Penrose Limits }

    The part of the supergravity Lagrangians that is relevant to the
study of AdS/CFT and DW/QFT dualities in various dimensions
\cite{tow}\cite{berg} can be written  as
\be {\cal L}_{D} = \frac{1}{2\kappa_{D}^{2}} e^{\delta \phi} [ - R
*\oneone + \tilde{\gamma} d\phi \wedge *d\phi +
\frac{1}{2}*F_{D-p-2} \wedge F_{D-p-2} ],
 \label{lag} \ee
where $\kappa_{D}$ is the gravitational coupling constant in D
dimensions and $\delta$, $\tilde{\gamma}$ are  two parameters that
will be specified below. The independent fields are the metric
$g_{MN}$, the dilaton $\phi$ and a $(D-p-3)$-form potential whose
field strength is $ F_{D-p-2} $. The integer $p$ corresponds to
the spatial dimension of the brane in the case of the p-brane
solutions, R is the scalar curvature of $g_{MN}$  and $*\oneone$
with the Hodge dual $*$ denotes the volume D-form. Our space-time
conventions and the field equations that follow from (\ref{lag})
are given in the Appendix.

     The parameters $\delta$ and $\tilde{\gamma}$ that appear in (\ref{lag}) are not
independent:
\be \delta = \frac{-(D-2)a}{2(D-p-3)}, \hspace{.5in}
\tilde{\gamma} = \frac{D-1}{D-2} {\delta}^{2} - \frac{4}{D-2},
 \ee
but  are determined in terms of D together with a choice of a
constant $a$ that specifies the relevant theory. For $D=11$
supergravity  $a = 0$ and the dilaton also vanishes. In $D=10$ one
finds that $a = (3-p)/ 2$ when $F_{D-p-2}$ is chosen to belong to
the RR sector of the type IIA or IIB supergravity theory and in
these cases (\ref{lag}) gives rise to the Dp-brane solutions. On
the other hand, when $F_{D-p-2}$ is in the $ D = 10 $
Neveu-Schwarz (NS) sector one has $ a = -1 $ for $ p=1 $ and $ a =
1 $ for $ p = 5$. Other values of $a$ that are relevant to the
lower dimensional supergravity theories can be found in
\cite{berg}.

   An interesting feature of (\ref{lag}) is that $g_{MN}$ is the dual frame
metric. The Einstein frame and the string frame Lagrangians can be
obtained from (\ref{lag}) by the conformal mappings:
\be g_{MN}^{Einstein}= e^{- [a  / (D-p-3)]\phi}  g_{MN},
\hspace{.5in} g_{MN}^{String} = e^{[4  / (D-2)]\phi}
g_{MN}^{Einstein} \label{frt} \ee
and these three frames, of course, coalesce in $D=11$. (Notice
that in (\ref{lag}) we are working with the magnetic potentials.
In the particular case D=11, p=2, it is  $F_{7} = *F_{4}$ that
appears as the field strength.) One should also keep in mind that,
when D=10, p=3 and $F_{5}$ is taken to be in the RR sector, the
field equations of (\ref{lag}) must be supplemented with the
self-duality condition: $ * F_{5} = F_{5} $.

    Consider now the solutions of (\ref{lag}) under the
assumption that the space-time $M_{D}$ is topologically and
metrically a product, $ M_{D} = M_{p+2} \times K $, of a
$(p+2)$-dimensional space-time $M_{p+2}$ and a
$(D-p-2)$-dimensional compact manifold  $K$. Notice that
metrically this is  a frame-dependent condition within the
conformal class of $g_{MN}$, and depending on the nature of
$\phi$, the metric may lose its product form in other frames. For
example, if one takes as $\phi$ the lift up of a function which is
defined either on $M_{p+2}$ or on $K$, then one ends up with
warped product metrics in other frames. If one assumes that the
dilaton is a constant, the product structure of $g_{MN}$ is, of
course, preserved in all the frames. In such cases one can readily
construct exact solutions by taking $F_{D-p-2}$ to be proportional
to the volume-form $ Vol(K)$ of $K$. This is, of course, the
well-known Freund-Rubin mechanism which reduces the field
equations to the requirement that
 $M_{p+2}$ and $K$  are both Einstein spaces. With this reduction
the cosmological constant of $M_{p+2}$  turns out to be
non-positive.

   In D=11 supergravity this procedure is known to give rise two distinct families which can
be conveniently displayed in terms of $ F_{4} $. The first family
is obtained by taking $ F_4 = \frac{3}{\it{l}}Vol(M_4)$ in which
case the field equations reduce to
\be ^{4}R_{\mu\nu} = -\frac{3}{\it{l}^2} g_{\mu\nu}, \hspace{.7in}
^{7}R_{mn}= \frac{3}{2\it{l}^2} g_{mn},
 \ee
and admit $AdS_{4} \times S^{7} $ as a distinguished solution. The
dual geometry $AdS_{7} \times S^{4}$ belongs to the second family
for which $ F_4 = \frac{6}{\it{l}}Vol(K)$ and this implies
 \be
^{7}R_{\mu\nu} = -\frac{6}{\it{l}^2} g_{\mu\nu},\hspace{.7in}
 ^{4}R_{mn}=
\frac{12}{\it{l}^2} g_{mn}.
 \ee
Here and in the sequel  Greek $(\mu,\nu,...)$ and Latin
$(m,n,...)$ indices refer to the bases of $M_{p+2}$ and $K$
respectively and $\it{l}$ is a real constant. The left
superscripts denote the dimensions of the spaces. In any dimension
$D$ the cosmological constant $\Lambda$ is defined to be $ \Lambda
= [ \epsilon (D-1)(D-2) ] / 2 \it{l}_{D}^{2} $ where $\it{l}_{D} $
is the corresponding real constant and $\epsilon =\pm 1$ is
introduced to allow both signs for $\Lambda$.

   A similar family is encountered in the specialization of
(\ref{lag}) to the $D=10$, $p=3$  Lagrangian where $F_{5}$ is the
self-dual RR field. After setting $\phi$ to a constant and taking
$ F_5 = \frac{2\sqrt{2}}{\it{l}} [Vol(M_5) - Vol(K)]$, all the
D=10 field equations are satisfied if
\be ^{5}R_{\mu\nu} = -\frac{4}{\it{l}^2} g_{\mu\nu},\hspace{.7in}
^{5}R_{mn}= \frac{4}{\it{l}^2} g_{mn} .
 \ee
The distinguished member of this family is $AdS_{5} \times S^{5}
$.

   Although all of these examples involve a constant dilaton and
are familiar from the near-horizon limits of non-dilatonic branes,
similar families exist even when $\phi$ is allowed to depend on
the coordinates of $M_{D}$, provided either $M_{p+2}$ or $K$ (but
not necessarily  $M_{D}$ itself) admits a conformal Killing
vector. This was first observed in \cite{duff} for the D=10, NS
$p=5$ Lagrangian where a CKV of $S^3$ was related to the gradient
of $\phi$ and a family which includes the $AdS_{7} \times S^{3}$
solution was obtained. In a similar analysis for the D=10, NS
$p=1$ Lagrangian, $AdS_{3} \times S^{7}$ type of solutions were
found by utilizing a CKV of $M_3$ which gave an invariant
interpretation of the  "{linear dilaton}"\cite{gib}.

     Association of $\phi$ with a conformal Killing potential
can be extended to the $D=10$, $p>3$ Lagrangians when $ F_{D-p-2}
$ is taken to be a RR field strength. When $p=4$, letting $ F_4 =
\frac{6}{\it{l}}Vol(K)$  and demanding at the same time that
$\psi= e^{2\phi/3}$ is a conformal Killing potential on $M_{6}$:
\be \nabla_{\mu} \nabla_{\nu} \psi = -\frac{1}{\it{l}^2} \psi
g_{\mu\nu}, \label{cke}\ee
 gives

\be ^{6}R_{\mu\nu} = -\frac{5}{\it{l}^2} g_{\mu\nu}, \hspace{.7in}
^{4}R_{mn}= \frac{12}{\it{l}^2} g_{mn}.
 \ee
The near-horizon $AdS_{6} \times S^{4}$ geometry of the D4-brane
is in this family.

  Some interesting changes occur when one moves to higher values
of p in the Dp-brane type Lagrangian. By letting
$\psi=e^{3\phi/4}$ for p = 5, it can be inferred that, when
$F_{3}$ is chosen on $K$ in the above manner,  $\nabla_{\mu}\psi$
must be a Killing vector on $M_{7}$: $ \nabla_{\mu} \nabla_{\nu}
\psi =0 $. This gives a Ricci-flat $M_{7}$ while keeping $K$ as an
Einstein space. On the other hand, in the p=6 case one finds that,
if $ F_{2} =\frac{2}{\it{l}}Vol(K)$ and
 \be
^{8}R_{\mu\nu} = -\frac{7}{\it{l}^2} g_{\mu\nu},\hspace{.7in}
 ^{2}R_{mn}=
\frac{4}{\it{l}^2} g_{mn}, \label{2a}
 \ee
then all field equations will be satisfied provided $\psi=
e^{-2\phi/3}$ is a solution of (\ref{cke}) on $M_{8}$ that also
obeys the condition
\be \nabla_{\mu}\psi \nabla^{\mu}\psi = -\frac{1}{\it{l}^2}
\psi^2.\label{sl}
 \ee
Hence in this case one is forced on a particular CKV which must be
\emph{space-like}. The $AdS_{8} \times S^{2}$ near horizon limit
of the D6-branes is in this category\footnote{ It may be useful to
note that the variable $ \Phi= e^{\delta\phi}$, which is used in
the field equations in the Appendix, is related to $\psi$ by
$\psi= \Phi, {\Phi}^{3/8}, {\Phi}^{-1/9}$ for $ p= 4,5,6 $
respectively.}.

    Suppose now we wish to take the Penrose limit of such an $M_{D}$ using
the D-dimensional scaling rules \cite{guv}. Recall that the
Penrose limit is a local procedure which requires first the
introduction of the appropriate coordinates in a conjugate
point-free portion of a null geodesic congruence of $M_{D}$. Since
$M_{D}$ has the form $M_{D} = M_{d} \times K $ and the dual frame
line element $ d{s_{D}}^{2} = d{s_{d}}^{2} + d{s_{K}}^{2} $ where
$d = p+2$ and $d{s_{K}}^{2}$ is the line element of $K$, the set
of all null geodesics of $M_{D}$ splits into a union of two
disjoint subsets. Because $K$ has a Riemannian metric, the first
subset consists of the time-like geodesics of $M_{d}$ together
with the geodesics of $K$ and both of these are parametrized by
the same D-dimensional affine parameter.  (On the underlying
spaces these are, of course, not unit-speed curves.) In the second
subset one has the null geodesics of $M_{d}$ that are passing from
fixed points of $K$. In this case the affine parameter of $M_{d}$
coincides with that of $M_{D}$. It turns out that the outcome of
the Penrose limit depends crucially on the subset that contains
the chosen null congruence.

   For a congruence that belongs to the first subset, the Penrose
coordinates of $M_{D}$ can be constructed by choosing a
synchronous coordinates system\cite{pen2} on $M_{d}$ together with
a set of geodesic coordinates on $K$. The two null coordinates of
the Penrose patch are then defined by the sum and the difference
of two coordinates which measure the proper time on the time-like
geodesic and the length of the Riemannian geodesic. Using the
scaling rules for the supergravity fields it is then easy to see
that the limit of $M_{D}$ is a plane wave space-time with a
\emph{non-zero flux} $F_{D-d}$ and $ \phi$. The well-known plane
wave limits of the near horizon geometries of the non-dilatonic
branes\cite{bla} are  in this category.

A completely different situation arises when the congruence is
chosen from the second subset. In this case the usual Penrose
limit cannot result in a solution that has a non-zero flux. This
is because, the Penrose coordinates of $M_{D}$ consists of the
Penrose coordinates of $M_{d}$ together with a set of suitable
coordinates on K and the limit of the product $M_{d}\times K $ can
be seen to be the product of the separate limits of $M_{d}$ and $
K $. The two null coordinates of the Penrose patch are now on
$M_{d}$ and since $F_{D-d}$ is a $(D-d)$-form on K, it picks up
the $(D-d)$ powers of the scaling parameter from the coordinates
of K. However, according to the supergravity scaling rules,  it
should have scaled with a power which is one less than its degree
in order to survive in the limit. Therefore, $F_{D-d}$ goes to
zero in the standard limit. This was observed for  $AdS_{p+2}
\times S^{D-p-2} $  in\cite{ff} and its limit was found to be the
$D$-dimensional Minkowski space.

The vanishing of the flux within the second subset is related  to
the fact that the Penrose procedure cannot give rise to a non-zero
cosmological constant on the limit of $M_{d}$. Moreover, the
standard Penrose limit does not take into account the possibility
of having a metric that contains  functions homogenous of degree
zero in the coordinates. Let us assume that K admits a coordinate
system in which its metric is a homogeneous function of degree
zero in $\it{l}$ and of these coordinates. For example, let us
take $ K=S^{D-d}$ with the standard metric and adopt the
stereographic coordinates of $S^{D-d}$ as part of the Penrose
patch of $M_{D}$. On the other hand, let us assume that $M_{d}$
admits a space-like CKV. Under these assumptions a generalization
of the Penrose limit which allows a non-zero cosmological constant
on $M_{d}$ as well as a non-zero flux and involves also a scaling
of $\it{l}$ is available\cite{guv2}. Let $\Omega$ be the real
scaling parameter. Since one of the scaling rules of\cite{guv2} is
$ \it{l} \rightarrow  \Omega \it{l}$, it can be checked that
$S^{D-d}$ and $F_{D-d}$ are not affected by this generalized
limit. The corresponding limit $M^{(0)}_{d}$ of $M_{d}$ was found
to be a generalized AdS plane wave space-time\cite{guv2}.
Therefore, when one chooses a null congruence that belongs to the
second subset and takes $K=S^{D-d}$, the $D$-dimensional solution
$M_{D}$ will have a limit which is the product of a
$d$-dimensional AdS plane wave and $S^{D-d}$ in the dual frame.
This limit is also an exact solution of the field equations with a
non-zero flux and will be called the conformal Penrose limit of
$M_{D}$. Notice that in the particular case $M_{D} = AdS_{d}
\times S^{D-d}$ the solution is mapped into itself under the
conformal Penrose limit.

\section{The Conformal Penrose Limit}

    As in the above $K = S^{D-d}$ example, we shall assume in
general that $K$ and $F_{D-d}$ are not affected by the conformal
Penrose limit and examine in detail its action on $M_{d}$. Noting
also that the only space-time which admits a space-like CKV and
has $\Lambda < 0 $ in $d=4$ is the $AdS_{4}$ itself\cite{gar}, we
shall concentrate on the cases $d > 4$.

The implications of the existence of a CKV field on $M_{d}$ is a
well-studied subject, dating back to the classical work of
Brinkmann \cite{br}, although the global results are quite recent
for the Lorentzian metrics, see e. g.
\cite{kt}\cite{ker}\cite{knel }\cite{bg }. When $M_{d}$ is an
Einstein space:
\be
R_{\mu\nu} = [\epsilon(d-1)/\it{l}^2]g_{\mu\nu}, \label{ein} \ee
and a smooth vector field $ V^{\mu} $ satisfying
\be {\mathcal{L}}_V g_{\mu\nu}=2\psi g_{\mu\nu},
\ee
exists, where $\mathcal{L}$ is the Lie derivative and $ \psi$ is a
smooth function on $M_{d}$, it can be deduced that $\nabla_{\mu}
\psi$ must itself be a CKV
\be \nabla_{\mu} \nabla_{\nu} \psi =  \frac{\epsilon}{\it{l}^2}
\psi g_{\mu\nu}, \label{ck2}\ee
(we shall assume $\psi \neq const.$ to exclude the homotheties)
and that
\be
 V_{\mu} = \epsilon \it{l}^2 ( \xi_{\mu} + \nabla_{\mu}
\psi ),\label{kp}\ee
where  $\xi_{\mu}$ is an ordinary Killing vector.
 The
algebra of conformal vector fields on $M_{d}$ therefore decomposes
into the direct sum of the algebra of the Killing vectors and the
algebra of closed CKV's that are locally gradients.

     An interesting distinction between the Riemannian and Lorentzian
metrics is the number of the fixed points of $V^{\mu}$ and these
points play an important role globally\cite{knel}\cite{bg}. Since
each  $V^{\mu}$ gives rise to a closed CKV, it is convenient to
concentrate on the fixed points of $\nabla_{\mu} \psi$. These are
the critical points of $\psi$ which can be shown to be isolated
points\cite{knel}. Around any point with $\nabla_{\mu}\psi
\nabla^{\mu}\psi \neq 0$ one can find a neighborhood where
$g_{\mu\nu}$ is a warped product metric and in this neighborhood a
coordinate system $ \{ y, x^{a}\}, a= 1, ...,(d-1)$, exists where
$\nabla_{\mu} \psi = U(y) \delta^{y}_{\mu}$,  $U= d\psi /dy$ and
\be
d{s_{d}}^2 = \eta dy^2 + U^{2}(y) g_{ab}(x) dx^{a}dx^{b}. \label{wm}
\ee
Here $g_{ab}(x)$ is a metric on the (d-1)-dimensional fibers and
$\eta = \pm 1$ is chosen according to the type of the CKV:
$\nabla_{\mu}\psi \nabla^{\mu}\psi = \eta U^2$. Notice that
whereas $\epsilon$ is the sign of the cosmological constant,
$\eta$ is the sign of the pseudo-norm of
 $\nabla_{\mu} \psi$ and these quantities  are, of course,
 independent. For a space-like CKV  one has $\eta = -1$ for
 any sign of $\Lambda$. It is also useful to note  that the critical points of $\psi$
 are now located at the zeros of U where the coordinate system
 breaks down.

    For (\ref{wm}) the field equations (\ref{ein}) require that
\be
U^{''}= \frac{\epsilon \eta}{\it{l}^2} U,
\ee
and consequently,
\bea
U = A\, cosh(y/\it{l}) + B\, sinh(y/\it{l}), \hspace{.2in}( \epsilon = \eta), & \nonumber
\\
U = A\, cos(y/\it{l}) + B\, sin(y/\it{l}), \hspace{.2in}( \epsilon =- \eta), &
\eea
where A and B are real constants and a prime denotes
differentiation with respect to $y$.  Moreover, according to
(\ref{ein}) the fiber metric $g_{ab}(x)$ must also be Einstein:
\bea
^{(d-1)}R_{ab} = [ \epsilon (d-2)(A^2 - B^2)/ \it{l}^2 ] g_{ab}, \hspace{.2in}( \epsilon = \eta), & \nonumber
\\
^{(d-1)}R_{ab} = [ \epsilon (d-2)(A^2 + B^2)/ \it{l}^2 ] g_{ab}, \hspace{.2in}( \epsilon =- \eta), &
\label{neq}\eea
which now guarantees that (\ref{ein}) is fully satisfied.
Choosing $ \epsilon = \eta = -1 $, $A=-B=1$ and
taking  $g_{ab}(x)$ as the (d-1)-dimensional flat Minkowski metric
gives $ AdS_d $ in Poincar$\acute{e}$ coordinates.
In this case the zero of U is located at the AdS horizon.

    According to (\ref{wm}), $M_{d}$ is a warped product:
$M{_d} = I \times_{U} N$, where $I$ is a real interval and  the
fiber $N$ is a $(d-1)$-dimensional manifold, and the structure of
the set of all null geodesics of $M{_d}$ as well as the nature of
$N$ depend on $\eta$ in an obvious manner. If $\eta = -1$, $N$ is
a Lorentzian manifold and the set of all null geodesics of $M{_d}$
is again a union of two disjoint subsets. The first subset is
composed of the null geodesics along which $y$ is not constant and
for these $\psi(y)$ is an affine parameter. The second subset
consists of the null geodesics of $N$ that are passing from fixed
points of $I$ so that $y = const.$ for these geodesics. In both
subsets the points $U = 0$ are conjugate points for the null
geodesics. Notice that when $\eta = 1$ so that the CKV is
time-like, $N$ must be a Riemannian manifold and then the second
subset is  obviously not available.

    Before considering the conformal Penrose limits relative to these subsets
let us momentarily specialize to the  $\epsilon = \eta =-1$
solutions which are our main concern, and note that such $M_{d}$
are conformally compactifiable provided $ U > 0$. The conformal
boundary corresponds to $ U \rightarrow \infty$ and the inverse of
U is a defining function. Letting $r=U^{-1}$, one can find a
manifold $\bar{M}$ with boundary $\partial \bar{M}= N $ such that
$M_{d}$ is diffeomorphic to $ \bar{M} -  \partial \bar{M}$. The
Einstein metric $\bar{g}_{\mu\nu} $ of  $\bar{M}$ is $
\bar{g}_{\mu\nu} = r^{2}g_{\mu\nu}$ and on $\partial \bar{M}$,
$r=0$  but $ \bar{g}^{\mu\nu}\nabla_{\mu} r \nabla_{\nu}r =
-\it{l}^{-2} $. In general $\partial \bar{M} $ is a time-like
boundary and may not be connected. Since $\partial \bar{M}$ need
not be topologically $ {\mathbb{R}} \times S^{(d-2)}$ and
$r^{(4-d)}\bar{C}_{\mu\nu\kappa\sigma}$ need not vanish at the
boundary, in general $M_{d}$ will not be asymptotically $AdS_{d}$.
Here $\bar{C}_{\mu\nu\kappa\sigma}$ is the Weyl tensor of
$\bar{M}_{d}$. In fact assuming $\Lambda < 0 $, it is easy to see
that the only $M_{d}$ which admits a space-like closed CKV and is
asymptotically $AdS_{d}$ is the $AdS_{d}$ space itself. Hence
$AdS_{d}$ can be uniquely singled out  by imposing the appropriate
boundary conditions together with the presence of a CKV when $d >
4$.

Returning back to (\ref{kp}), let us write the Killing vector as
$\xi^{\mu} = (\beta, \xi^{a})$. The Killing equation then implies
that $\beta$ must be a scalar field on $N$ and if we define   $
\xi_{a} = g_{ab}\xi^{b} $ with the decomposition $\xi_{a} =
f(y)\nabla_{a} \beta(x) + \zeta_{a} (x) $ it follows that either
$\nabla_{a}\beta =0$ or $U^2 f'(y) = -\eta$. In the latter case
$f(y)$ can be easily determined and $ \nabla_{a} \beta $ can be
seen to be a non-homothetic CKV of $N$. Then $\zeta_{a}$ is an
ordinary Killing vector on $N$. On the other hand, if
$\nabla_{a}\beta =0$, one can show that $\beta = 0$ as long as
$U'/U \ne const.$ and $\zeta_{a}$ is again a Killing vector on
$N$. In the particular case $U'/U = const.$, a non-zero constant
$\beta$ can exist and $\zeta_{a}$ becomes a homothetic Killing
vector on $N$. In this manner the isometries of $M_{d}$ give rise
to the conformal motions on the boundary $N$.

Consider now the conformal Penrose limit of $M_{d}$ for all values
of $\epsilon$ and $\eta$. First let us suppose that $\epsilon =
\eta =-1$ and $M_{d}$ admits a  Killing vector $\xi^{\mu}$
associated with $V^\mu$. If one is then interested in setting up
the Penrose coordinates around a congruence that belongs to the
first subset, the coordinate system of (\ref{wm}) is not a
convenient starting point. In this case the procedure described
in\cite{guv2} can be taken over and a neighborhood in which
$V^\mu$ has no fixed points can be blown up to get
\be
 d{s_{d}}^2 = \frac{\it{l}^2}{[z/\lambda +b_k(u)x^k + \it{l}
c(u)]^2}[2 du dv - h_{ij}(u) x^i x^j du^2 - \delta_{ij}dx^i dx^j
-(dz - \gamma du)^2],\label{met} \ee
as the conformal Penrose limit of $M_{d}$. Here  $x^{i},
i=1,...,d-3,$ is a set of $(d-3)$ transverse coordinates, $\gamma
= -2\lambda \dot{b}_{j}x^j + \dot{\lambda} z/\lambda$ and due to
(\ref{ein}), the metric functions satisfy
\be \lambda^{-2} + b_{j}b_{j} =1, \label{con} \ee \be \ddot{c} =
0, \ee \be \ddot{b}_{j} = h_{jk}\:b_k, \ee and \be h_{jj}=  -
\ddot{\lambda}/\lambda - 2{\lambda}^2 \dot{b}_j \dot{b}_j , \ee
where a dot denotes differentiation with respect to the null
coordinate $u$ and repeated indices are summed with $\delta_{ij}$.
This is just the metric of a generalized AdS plane wave for which
$V^{\mu} = \lambda(u) \delta^{\mu}_{z}$ is the CKV inherited from
$M_{d}$. The vector $ b_{j}(u)$ characterizes the twist of
$V^{\mu}$; if $\dot{b}_{j}(u)= 0$,   $V^{\mu}$ is hypersurface
orthogonal in which case it is reducible to a closed vector field.
Another useful representation of (\ref{met}),  in terms of a new
coordinate $\tilde z$, is
\be
 d{s_{d}}^2 = \frac{\it{l}^2}{{\tilde z}^2}[2 du dv - h_{ij}(u) x^i x^j du^2 - \delta_{ij}dx^i dx^j
-\lambda^{2} (d{\tilde z} + {\mathcal{A}})^2],\label{met1} \ee
where the Kaluza-Klein (KK) gauge field is given by $
{\mathcal{A}} = (\dot{b}_{j} x^{j} - \it{l} \dot{c}) du - b_{j}
dx^j$.

It should be noted that the conformal Penrose limit of $M_d$
always has the metric (\ref{met}) whenever $\eta = -1$ for the
above choice of the null congruence and the CKV. However, when the
cosmological constant of $ M_d$ is taken to be positive,
(\ref{ein})  requires in the limit in place of (\ref{con}) that
\be
 \lambda^{-2} + b_{j}b_{j} = -1,
\ee
and it follows that $M_d$ does not have a conformal Penrose limit
if $\eta = -1, \epsilon = 1$ and the congruence is chosen from the
first subset.

 On the other hand, if one  would start from the premise
that $\eta = 1$, but $\epsilon= \pm 1$, then the gauge choice on
$M_{d}$ which leads to (\ref{met}) would not be
available\cite{guv2}. Hence such an $M_{d}$ would not have any
conformal Penrose limit because the second subset of the null
geodesics would also be absent. We may therefore conclude that
$M_d$ must admit a space-like CKV in order to have a well-defined
conformal Penrose limit.

For a null geodesic congruence that belongs to the second subset
when $\eta = -1$, the Penrose coordinates of $M_{d}$ can be taken
to be the  Penrose coordinates of $N$ together with the coordinate
$y$. Consequently, the limit involves the standard scalings of the
Penrose coordinates of $N$ together with the scalings: $y
\rightarrow \Omega y$ and $ \it{l} \rightarrow  \Omega \it{l}$
which leave U invariant.  This means that on $N$ the procedure
will amount to the standard Penrose limit which will force the
limit of $N$ to be a Ricci flat, ordinary plane wave space-time.
Since one would like to keep the $d$-dimensional $\Lambda \neq 0 $
in the limit and (\ref{neq}) must hold, this will only be possible
if $\epsilon = \eta$ and $A^2 = B^2$. When these hold one gets as
the limit
\be
 d{s_{d}}^2 = \frac{\it{l}^2}{z^2}[2 du dv - h_{ij}(u) x^i x^j du^2 - \delta_{ij}dx^i dx^j
- dz^2],\label{met2} \ee
which can be interpreted as the non-linear version of the Randall-Sundrum zero mode\cite{chg}
and corresponds to the $\dot{b}_j =0$ specialization of (\ref{met}).

The above discussion which led us to the limit (\ref{met2}), of
course, disregards once again the possible existence of a
space-like CKV, this time on $N$. If $N$ admits such a CKV, then
the above arguments can be applied recursively to $N$ in which
case $N$ will again have a warped product structure.

Notice also that so far we have  considered only the neighborhoods
of $M_{d}$ without fixed points. It is known\cite{knel} that in a
neighborhood which contains a critical point of $\psi$, $M_{d}$ is
isometric to the Poincar$\acute{e}$ patch of $AdS_{d}$ and
consequently, this neighborhood of $M_{d}$ will be invariant under
the conformal Penrose limit. The remaining possibility is the case
of a $\nabla_{\mu} \psi$ that is null on a neighborhood of $M_{d}
$. In this case $M_{d}$ is already a Ricci-flat pp-wave
space-time\cite{br}.

\section{Properties of the Limiting Solution}
We have thus seen that $M_d$ has a conformal Penrose limit only
when its cosmological constant is negative and the CKV is
space-like. Since it is precisely these type of $M_d$ that solve
the supergravity equations in the dual frame, we now examine the
properties of the limiting solution $M^{(0)}_{d}$. In the
particular case $\dot{b}_{j}(u)= 0$ it is already known that
$M^{(0)}_{d}$ preserves $1/4$ of the maximal
supersymmetry\cite{bre} and has a Virasoro symmetry\cite{ban}. Our
first goal is to see whether these properties can be extended to
the general solution (\ref{met}) which allows a non-zero twist.

We start by considering the existence of the Killing spinors that
satisfy
\be D_{\mu} \varepsilon - \frac{i}{2\it{l}} \Gamma_{\mu}
\varepsilon = 0, \label{ks}\ee
where $D_{\mu}$ is the spinor covariant derivative and
$\Gamma_{\mu}$ are the Dirac matrices on $M^{(0)}_{d}$. In the
coordinate system of (\ref{met}) it will be convenient to let $W =
[z/\lambda +b_k(u)x^k + \it{l} c(u)]$ and choose the orthonormal
basis one-forms as
\bea
 {\bf{e}}^{0} = \frac{\it{l}}{\sqrt{2} W} \{ dv + \gamma dz + [1 -
\frac{1}{2}(h_{ij}x^i x^j + \gamma^2)] du \}, & \nonumber \\
{\bf{e}}^{1} = \frac{\it{l}}{\sqrt{2} W} \{ dv + \gamma dz - [1 +
\frac{1}{2}(h_{ij}x^i x^j + \gamma^2)] du \}, & \nonumber \\
{\bf{e}}^j = \frac{\it{l}}{W} d x^j, \hspace{.1in}  (j=1,...,d-3),
\hspace{.4in} {\bf{e}}^d = \frac{\it{l}}{W} dz. \eea
If one then defines
\be \Gamma_{\pm} = \frac{1}{\sqrt{2}} (\Gamma_{0} \pm \Gamma_{1}),
\ee 
and
\be \varepsilon = \sqrt{\it{l} /W} {\bar{\varepsilon}},
\label{cs}\ee
the integrability condition for (\ref{ks}) can be seen to be 
\be [ (h_{jk} +3 {\lambda^2} \dot{b}_{j} \dot{b}_{k}) \Gamma^k -
(\lambda \ddot{b}_{j} + 3 \dot{\lambda} \dot{b}_{j} )\Gamma^d ]
\Gamma_{+}{\bar{\varepsilon}} = 0. \label{ic} \ee 
Therefore, whenever $h_{jk} \ne 0$ and the matrix within the
square brackets is invertible, the necessary condition for the
existence of a Killing spinor is 
\be \Gamma_{+}{\bar{\varepsilon}} = 0.\label{hss} \ee 
The vanishing of $h_{jk}$ implies that $\dot{b}_j = 0$ in which
case  $M^{(0)}_{d} = AdS_d $ and (\ref{ic}) is identically
satisfied. On the other hand, if one imposes rather than
(\ref{hss}) the condition: 
 \be [ (h_{jk} +3 {\lambda^2} \dot{b}_{j} \dot{b}_{k}) \Gamma^k -
(\lambda \ddot{b}_{j} + 3 \dot{\lambda} \dot{b}_{j} )\Gamma^d ]
{\bar{\varepsilon}} = 0, \ee 
then it is easy to see that such a spinor does not allow a
$\dot{b}_j = 0$ specialization as long as  $h_{jk}$ is an
invertible $(d-3) \times (d-3)$ matrix. Because of this reason we
impose (\ref{hss}) as a necessary condition and its effect is to
eliminate 1/2 of the $AdS_d$ supersymmetry.

When (\ref{hss}) holds, the components of (\ref{ks}) can be cast into the form: 
\bea
\partial_{v}{\bar{\varepsilon}} = 0, & \nonumber \\
\partial_{j}{\bar{\varepsilon}} -\frac{1}{2W} \Gamma_j Q_{+}{\bar{\varepsilon}} =
0,& \nonumber \\
\partial_{z}{\bar{\varepsilon}} -\frac{1}{2W} \Gamma_d Q_{+}{\bar{\varepsilon}} =
0,& \nonumber \\
\partial_{u}{\bar{\varepsilon}} + \frac{1}{2} \lambda \dot{b}_j
\Gamma_j \Gamma_d {\bar{\varepsilon}} - \frac{1}{2W} \Gamma_{-}
Q_{+}{\bar{\varepsilon}} =0, \label{ks1} \eea where
$\partial_{\mu}$ are the partial derivatives with respect to the
coordinates and $Q_{\pm}$ are the
operators 
\be Q_{\pm} = \Gamma^{\mu}\partial_{\mu} W \pm i I, \ee
with $I$ denoting now the identity operator. These operators can
depend only on the coordinate $u$ and
satisfy 
\be Q_{\pm}^2 = \pm 2iQ_{\pm}, \hspace{.2in} Q_{+}Q_{-} =
Q_{-}Q_{+} =0, \ee as a consequence of (\ref{con}). One can impose
\be Q_{+} \bar{\varepsilon} = 0, \label{qs} \ee as an additional
algebraic condition on the Killing spinor and this can be ensured
by setting \be \bar{\varepsilon} = Q_{-}\alpha(u). \ee
 With these conditions Killing spinor equations now reduce
to an equation that just fixes the $u$-dependence of the spinor
$\alpha$: \be \frac{d}{du} [Q_{-}\alpha(u)] + \frac{1}{2} \lambda
\dot{b}_j \Gamma_{j} \Gamma_{d} Q_{-} \alpha(u) = 0. \label{ks2}
\ee Since (\ref{ks2}) always has a solution and (\ref{qs})
eliminates one half of the remaining supersymmetry we conclude
that $M^{(0)}_d$ possesses 1/4 of the maximal supersymmetry.
Notice that \be Q_{+} = i I - (\lambda^{-1} \Gamma_{d} +
b_{k}\Gamma_{k}), \ee and when $\dot{b}_j = 0$,  (\ref{qs})
becomes $\Gamma_{d} \bar{\varepsilon}= i \bar{\varepsilon}$ and
$\alpha$ reduces to a constant spinor.  Hence in this particular
case Killing spinors are given simply by (\ref{cs}) where
$\bar{\varepsilon}$ is a constant spinor that obeys (\ref{hss})
and (\ref{qs}).

We next display an infinite-dimensional symmetry of the metric
(\ref{met}) which generalizes the Virasoro symmetry of \cite{ban}.
Consider the diffeomorphism of $M^{(0)}_d$ defined by \bea
 u = f(U), \hspace{.2in} & \nonumber\\
x^j =\sqrt{f'}\; y^j, \hspace{.15in} & \nonumber\\
z =  \sqrt{f'}\; Z, \hspace{.15in} & \nonumber\\
v = V + {\frac{1}{4}} (y_k y_k + Z^2) \frac{f''}{f'},\label{tr}
\eea where $f$ is an arbitrary differentiable function and primes
denote the differentiations with respect to $U$. This
transformation maps (\ref{met}) to the metric \be
 ds^2 = {\it{l}}^{2} W_{new}^{-2} [ 2dU dV - H(U, y^j, Z) dU^2
 - \delta_{jk} dy^j dy^k -(dZ - \gamma (U) dU)^2 ], \ee
with the new metric functions: \bea W_{new} = Z/\lambda + b_{k}
y^{k} +
\it{l}\; e(U), \hspace{.2in} \gamma(U) = -2\lambda \; b'_{k} y^k +  \lambda ' \;Z / \lambda,& \nonumber \\
H(U, y^j,Z)= (f')^{2}\; h_{jk} y^j y^k - \frac{1}{2} (y_k y_k +
Z^2) \{ f, U\} - \frac{f''}{f'}\;\gamma(U) Z , \label{mf}\eea
where $ \{ f, U\}$ denotes the Schwarzian derivative: \be \{ f,
U\} = \frac{f'''}{f'} - \frac{3(f'')^2}{2(f')^2}, \ee and $e(U) =
c(u)/\sqrt{f'}$  satisfies \be e'' + \frac{1}{2} \{ f, U\} e = 0.
\ee The diffeomorphism (\ref{tr}) is precisely the one that was
utilized in \cite{ban} to derive the Virasoro symmetry and
(\ref{mf}) reduces to the corresponding transformation of the
metric function when $\dot{b}_j =0$.  In our context (\ref{tr}) is
a mapping within the family of generalized AdS plane waves.

 Finally let us consider the pp-curvature singularity of $M^{(0)}_d$.
Most of the $\epsilon = \eta = -1$ spaces $M_{d}$ that we had
studied in the previous section locally are not geodesically
complete. In fact one can show that if $M_{d}$ is a geodesically
complete Einstein space, then it is either $AdS_{d}$ or must have
the form $M_{d}=\mathbb{R}$ $ \times_{U} N$ where $ U = A
cosh(y/l)$ and $N$ is a complete Einstein space\cite{knel}. It is
easy to see that, unless $N$ is allowed to possess a space-like
CKV, geodesically complete $\mathbb{R}$ $ \times_{U} N$
space-times will not have a conformal Penrose limit. When $N$ is
assumed to have the desired CKV, taking the limit recursively
shows that the conformal Penrose limit of this particular
$\mathbb{R}$ $ \times_{U} N$ space-time is again itself. Hence the
$AdS$ plane waves  are obtained as the conformal Penrose limits of
$M_{d}$ that are not geodesically complete.

 In particular, the $A^2 = B^2$ subset of solutions which have
$U= A e^{(\pm y/{\it{l}})}$ and tend to (\ref{met2}) are not null
complete even when $N$ is complete\cite{on}. The limits
(\ref{met2}) of these solutions are known to have pp-curvature
singularities\cite{po}\cite{bre} at $z=\infty$ whenever $h_{ij}(u)
\ne 0$ and $d>3$. It can be checked, by examining the components
of the Riemann tensor in a frame that is parallelly transported
along a time-like geodesic, that the same behavior persists even
when the CKV vector is not hypersurface orthogonal. In other
words, the metric (\ref{met}) also has a pp-curvature singularity,
this time when $z/\lambda +b_k(u)x^k + \it{l} c(u)$ tends to
infinity. In the coordinates of (\ref{met1}), the pp-curvature
singularity is located at $ \tilde{z} = \infty$. We therefore see
that $M^{(0)}_{d}$ has two boundaries; there is a conformal
boundary $N$ which is positioned at  $ \tilde{z} = 0$ and which is
a perfectly well-behaved space-time with the metric:
 \bea
 ds^{2}_{N} = 2dudv - [h_{jk} x^j x^k + \lambda^2 \; (\dot{b}_j
x^j -\it{l} \dot{c})(\dot{b}_k x^k -\it{l} \dot{c})]\;du^2 +
2\lambda^2 \;(\dot{b}_j x^j -\it{l}\dot{c})\;b_k dx^k du &\nonumber \\
- (\delta_{jk} +\lambda^2 \; b_j b_k) \; dx^j dx^k. \hspace{3.5in}
\label{mets}
 \eea
This shows that in general $N$ is  a $(d-1)$-dimensional,
Ricci-flat pp-wave space-time with a special dependence on the
transverse coordinates. When $\dot{b}_{j} =0$, $N$ possesses the
standard plane wave metric in the harmonic coordinates. On the
other hand, $M^{(0)}_{d}$ also has a pp-curvature singularity  at
$ \tilde{z} = \infty$ which should be regarded as a physical
boundary when viewed from the $d$-dimensional perspective. Since
$M_{D}$ has the limit $M^{(0)}_{d}\times S^{D-d}$, it is clear
that $ \tilde{z} = \infty$ is also a genuine singularity from the
$D$-dimensional viewpoint. This raises the question whether these
pp-curvature singularities can be resolved by some means.

It is known that certain scalar polynomial singularities of the
Riemann tensor can be resolved when $ p=d-2$  is \emph{odd} by
viewing the same solution in a higher dimension\cite{ght}.
Remarkably, it turns out that the above pp-curvature singularities
of the $D=10$ solutions can be resolved only for a particular
\emph{even} value of $p$ when the solutions are lifted up to
$D=11$. This occurs when $p=6$ and the field equations of type IIA
supergravity theory reduce to (\ref{2a}). Recall that this is the
only case in which the CKV is forced to be space-like by the field
equations.

In the following let us distinguish the $D=11$ supergravity fields
with hats and note that the field content of (\ref{lag}) in the
case of $D=10$ type IIA supergravity implies for $p = 6$
\be
 d\hat{s}^2 =e^{4\phi/3} [ ds^{2}_{10}-  (d Y
+{\hat{A})}^2],\label{met11}
 \ee
and $\hat{F}_{4}=0$. Here $Y$ is the $D=11$ Killing coordinate
coordinate used in the reduction, $\hat{A}$ is the KK one-form
with  $F_{2}= d\hat{A}$ and $ds^{2}_{10}$ denotes the $D=10$ line
element in the dual frame. Since $\hat{F}_{4}=0$, the oxidation of
the family (\ref{2a}), which includes its conformal Penrose limit,
gives rise to Ricci-flat solutions of $D=11$ supergravity.

 Let us also recall that for the present $p=6$ case, $\psi =
e^{-2\phi/3}$ and  $e^{4\phi/3}= {\tilde{z}}^2$ in the coordinate
system of (\ref{met1}). According to (\ref{met11}) and
(\ref{met1}) the $D=11$ vacuum solution which is obtained by
oxidizing the metric of $M^{(0)}_{8}\times S^{2}$ is therefore
\be d{\hat{s}}^2 = {\it{l}^2}[2 du dv - h_{ij}(u) x^i x^j du^2 -
\delta_{ij}dx^i dx^j -\lambda^{2} (d{\tilde z} + {\mathcal{A}})^2
] -{\tilde{z}}^2 [ (d Y +\hat{A})^2 + d\Omega^{2}_{2}
],\label{m11} \ee
where $  d\Omega^{2}_{2} $ is the metric of a sphere of radius $
{\it{l}/2}$ and for the transverse coordinates $x^i$ one now has
$i=1, ... ,5$. Since $F_{2}=\frac{2}{\it{l}}Vol(S^{2})$, it
follows that the $D=11$ manifold  is  the warped product:
$\hat{M}={\tilde{M}}_{8} {\times}_{\tilde{z}^{2}} B_{3}$ where
$B_{3}$ is a $U(1)$ bundle over $S^2$ and ${\tilde{M}}_{8}$ is a
plane wave space-time. Using the spherical coordinates $( \theta,
\varphi )$ of $S^2$ and defining $\chi = 2Y/{\it{l}}$, one can
locally write
\be (d Y +\hat{A})^2 + d\Omega^{2}_{2}= \frac{\it{l}^2}{4}[
(d\chi+(1-cos\theta)d\varphi)^{2} + (d\theta^{2} + sin^{2}\theta
d\varphi^{2})]. \ee
If the coordinate $\chi$ on the $U(1)$ fibers is identified with a
period $4\pi$, $B_{3}= S^{3}$ and one has a Hopf fibration $S^3
\rightarrow S^2$. More generally, $B_{3}$ is a cyclic lens space.
When the gravitational degrees of freedom are switched off by
setting $h_{ij}(u) =0$, the solution (\ref{m11}) reduces to the
flat $D=11$ metric which was studied in\cite{im} as the lift up of
the near-horizon limit of the $D6$-branes.

We have  checked that all the components  $\hat{R}_{ABCD}$ of the
Riemann tensor, in a frame which is parallelly transported along a
time-like geodesics of (\ref{m11}), are regular everywhere
provided $\tilde{z} > 0$. In particular, the geodesics are
perfectly well-behaved as $\tilde{z} \rightarrow \infty$. We
therefore conclude that the pp-curvature singularity of
$M^{(0)}_{8}\times S^{2}$ is resolved in $D=11$ supergravity
theory.

\section{Discussion}

In this paper we have seen how the limiting procedure of
\cite{guv2} finds a natural application in the supergravity
theories as conformal Penrose limits of solutions in the dual
frame. For this purpose we have considered the solutions in which
the space-time is  of the form $M_{D} = M_{d} \times K$ where
$M_d$ is an Einstein space admitting a CKV and $K$ is a compact
Riemannian manifold. By studying the null geodesics in all
possible neigborhoods, we have seen that $M_D$ has a conformal
Penrose limit $M^{(0)}_{d} \times K$ if the cosmological constant
of $M_d$ is negative and the CKV is space-like. Under these
conditions $M^{(0)}_{d}$ was found to be a generalized $AdS$ plane
wave that possesses $1/4$ of the maximal supersymmetry as well as
a Virasoro symmetry.

 The basic requirement on $K$ was that it admits a coordinate
system in which its metric is a function homogeneous of degree
zero in $\it{l}$ and these coordinates. Although the limit was
seen to apply to any such $K$, we  have concentrated on the case $
K = S^{D-d} $ because this  choice had the virtue that $AdS_{d}
\times S^{D-d}$  is a  member of the family  and moreover, such a
coordinate system was explicitly known on $ S^{D-d} $. Since it is
also known that $S^{D-d}$ admits a CKV but this property was not
referred to in our discussion, it will be interesting to determine
the set of all $K$ that is left invariant by the conformal Penrose
limit.

We have also studied the global structure of $M^{(0)}_{d}$ and
noted that it has a conformal boundary $N$ which is located at $
\tilde{z} = 0$ and a pp-curvature singularity at  $ \tilde{z} =
\infty$. We have  found that in the $D=10$, $p=6$ case, the
pp-curvature singularity of $M^{(0)}_{8}$ can be resolved by
lifting up the solution to the $D=11$ supergravity theory. In this
case  one can verify  that (\ref{m11}) implies
$\hat{R}_{ABCD}\hat{R}^{ABCD} =0$ and $ \hat{M}$ can be viewed as
an interesting generalization of  the ordinary plane wave
space-times. What one now has in $D=11$ is an asymptotically
locally Euclidean (ALE) plane wave with an $A_{N-1}$
singularity\cite{ghaw} at $\tilde{z} = 0$ which depends on the
nature of the identifications on  $S^3$. Since we already know
that in the $D=10$ picture $\tilde{z} = 0$ defines the conformal
boundary  of $M^{(0)}_{8}$ and the boundary is a regular, $d=7$
pp-wave space-time, the  $A_{N-1}$ singularity is resolved in turn
in the corresponding eight-dimensional theory.

When the type IIA supergravity is compactified on $S^2$ or the
$D=11$ supergravity is compactified on a $U(1)$ bundle over $S^2$,
one obtains the $SU(2)$ gauged  $d=8$ supergravity theory of Salam
and Sezgin\cite{ss}.  The domain wall solution of this  theory was
constructed in\cite{lpt} and is known\cite{tow} to correspond to
the $h_{ij}(u) =0$ specialization of $M^{(0)}_{8}$. Even when
$h_{ij}(u) \neq 0$, the metric (\ref{met}) and the dilaton of
$M^{(0)}_{8}$ constitute a solution of the gravity/dilaton sector
of the $ SU(2)$ gauged theory. Hence the  $A_{N-1}$ singularity is
resolved in the $SU(2)$ gauged  $d=8$ supergravity.

Notice that the $SU(2)$ gauged $d=8$ supergravity theory has a
consistent brane-world KK reduction to $d=7$ ungauged
supergravity\cite{clp} and the pp-wave metric of the boundary $N$
is a Ricci-flat solution of this $d=7$ supergravity theory. This
raises the question whether the ungauged $d=7$ supergravity on the
pp-wave background $N$ can be in some sense dual to the $ SU(2)$
gauged supergravity theory on $M^{(0)}_{8}$. Since the $D6$-brane
worldvolume theory does not decouple from the bulk\cite{im}, it is
not unreasonable to expect a generalized sort of a duality between
two supergravities in the present context and an attractive
feature of the plane wave space-times is that they have a very
restricted set of quantum corrections\cite{gpl}\cite{des}. It will
therefore be an interesting problem to see whether such a duality
can be established between these two theories with the aid of the
ALE plane waves of the $D=11$ supergravity theory.

\section*{Acknowledgements}

I am grateful to M.J. Duff, G.W. Gibbons, C.N. Pope and E. Sezgin
for discussions and thank The Abdus Salam International Centre for
Theoretical Physics and Michigan Center for Theoretical Physics
for hospitality during the course of this work. The research
reported in this paper has been supported in part by the Turkish
Academy of Sciences (TUBA).

\section*{Appendix}
Our conventions are as follows:

 In all $D \geq 2$ we use the
``mostly minus'' signature $ (+,-, \ldots, -)$ and the orientation
$ \epsilon_{012 \ldots D-1} = 1$. The Ricci tensor is defined as
$R_{MN} = {R^L}_{MNL}$ and the Riemann
curvature obeys $ (\nabla_{N} \nabla_{M} - \nabla_{M} \nabla_{N})
T_K = {R^L}_{KMN} T_{L} $ for an
arbitrary $ T_M $. The Hodge dual of a p-form $(p \leq D)$ is
defined by
\begin{eqnarray}
*( W^{A_1} \wedge \ldots \wedge W^{A_p}) =
\frac{(-1)^{(D-1)}}{(D-p) !}
   \epsilon^{A_1 \ldots A_p A_{p+1} \ldots A_ D}
 W_{A_{p+1}} \wedge \ldots \wedge W_{A_D}, \nonumber
\end{eqnarray}
in terms of an orthonormal basis $\{ W^A \}$.

    The field equation that follow from (\ref{lag}) are
\bea
 d( *\Phi F_{D-p-2} ) = 0,\nonumber
\eea
\bea \triangle \Phi +
\frac{\delta^{2}(D-p-3)(-1)^{p(D-p)}}{8(D-p-2)!} {F^{2}}
 \Phi = 0, \nonumber\eea
\bea \lefteqn{ R _{MN} = -\Phi^{-1} \nabla_{M} \nabla_{N}\Phi +
{\tilde{\gamma}}
\delta^{-2} \Phi^{-2} \nabla_{M} \Phi \nabla_{N} \Phi} \nonumber \\
&+\frac{(-1)^{p(D-p)}}{2(D-p-3)!} F_{A_{1}... A_{D-p-3} M}
{F^{A_{1}... A_{D-p-3}}}_{N} -
\frac{(-1)^{p(D-p)}(4-\delta^{2})(D-p-3)}{8(D-2)(D-p-2)!} F^{2}
g_{MN},\nonumber \eea
where  $ \Phi= e^{\delta\phi}$ and
$F^{2} = F_{A_{1}... A_{D-p-2}}F^{A_{1}... A_{D-p-2}}$.

\end{document}